\documentclass[twocolumn,english,preprintnumbers,showpacs,showkeys,prl]{revtex4}

\usepackage{amsmath,amsfonts,amssymb}
\usepackage{graphicx}
\usepackage[colorlinks=true, allcolors=blue]{hyperref}

\usepackage{bm}
\usepackage{mathrsfs}
\usepackage{verbatim}
\usepackage{bbold}

\begin{document} 


\title{Higgs boson mass from  maximally nonlinear superconductive quantum gravity}

\author{Jeffrey Yepez}\email{yepez@hawaii.edu}
\date{July 31, 2019}
 \affiliation{
  Department\;of\;Physics\;and\;Astronomy,  University\;of\;Hawai`i\;at\;M\=anoa, Watanabe\;Hall,\;2505\;Correa\;Road, Honolulu,\;Hawai`i\;96822
}

\begin{abstract}
 Presented is a quantum gravity theory that is a quantum mechanical generalization of Einstein's vierbein field-based approach, where the classical metric tensor field is promoted to a quantum mechanical metric tensor field operator. 
The quantum gravity theory derives from quantum information dynamics intrinsic to quantized space, which is taken to be a tensor product space on a qubit array.  Hence, the metric tensor field operator is expressed as a product of two frame 4-vectors, which are anticommuting operators and naturally represented by Dirac matrices. The quantum gravity theory reduces to an effective nonlinear theory for a superconductive Fermi condensate.  The  asymmetric part of the metric tensor field operator encodes a fermion's intrinsic spin and mass in the torsion of space. 
A lower bound on the Fermi condensate's pair mass is found and  the pair's mass estimated.
\end{abstract}


\maketitle


{\it Introduction.}---With the hope of finding a renormalizable quantum gravity theory, one possibility is to search for an equivalent quantum computing model which encodes unitary particle and field dynamics using a quantum generalization of the classical metric tensor field of a curved-space manifold. Such an approach would be a modern realization of Mie's idea of an ``unavoidable connection" between gravitation and the existence of the fundamental particles \cite{Mie_1912a,Mie_1912b,Mie_1913}.  Moreover, if the quantum generalization of the metric tensor had an antisymmetric part encoding torsion of space, a possibility 
  anticipated by Sciama  \cite{RevModPhys.36.463}, then the intrinsic spin of a quantum particle becomes a gravitational phenomenon. The   4-vector basis of the local tangent space of the spacetime manifold in classical gravity could be generalized to a quantum mechanical 4-vector basis in the quantum gravity theory.

The idea that space by itself can support all particle and field dynamics dates back to Kaluza \cite{Kaluza_1921}, Einstein \cite{einstein-1928b}  and Wheeler \cite{PhysRev.97.511,misner-73}, leading to string theory \cite{Green_Schwarz_Witten_1987_v1,Green_Schwarz_Witten_1987_v2} based only on a gravitational Lagrangian as a theory of everything.  Recently, Kempf has explored the idea that space can be simultaneously discrete and continuous \cite{PhysRevLett.103.231301,PhysRevLett.92.221301,PhysRevLett.85.2873,PhysRevD.71.023503,PhysRevD.71.023503,PhysRevD.69.124014,1367-2630-12-11-115001}.  Also, the idea that curved space is expressible in terms of spatial entanglement offers a quantum computational opportunity to emulate particle and field dynamics with a lattice model equivalent to a curved-space theory \cite{yepez:770202,Kempf2014,PhysRevLett.116.201101}.

The quest for a unified field theory began in 1928 with EinsteinÕs vierbein field theory of gravity and electromagnetism. Although EinsteinÕs approach was not accepted at the time (see Born's July 15, 1925 letter \cite{Born_Einstein_letters}), today it is foundational to studies of quantum gravity  \cite{Birrell_Davies_82}.  Presented is a quantum gravity theory that is a quantum mechanical generalization of EinsteinÕs vierbein field-based approach. In particular, the  quantum gravity theory  is the dual of a traced-scaled vierbein-based gravity theory.
The quantum gravity theory presented here can yield quantitative predictions of Standard Model parameters---the mass of the Higgs boson in the Standard Model is provided as an example of this tractability.

{\it General relativity.}---Einstein's theory of classical General Relativity can be formulated as a gauge-gravity theory by using four vierbein fields \cite{einstein-1928a,einstein-1928b,yepez_arXiv1106.2037_gr_qc}.   
  In  Einstein's vierbein field formulation of  General Relativity, the basis vectors of the tangent space of a spacetime manifold do not derive from any coordinate system of that manifold.  Instead the basis vectors are a fixed noncoordinate basis. 
 The vierbein fields are a set of four Lorentz 4-vector fields---herein denoted by ${e^\mu}_a(x)$,  where $\mu=0,1,2,3$ is the contravariant 4-vector index in the coordinate (Greek) basis while $a=0,1,2,3$ is the covariant index in the noncoordinate (Latin) basis.   
Taken together, the set of vierbein fields ${e^\mu}_a(x)$ constitutes a transformation matrix (with 16 components) that  transforms a quantity, such as a 4-vector $V^a$, from the noncoordinate frame to  the coordinate frame,  $V^\mu(x) = {e^\mu}_a(x) V^a$.  

Einstein's unified field theory is a classical theory of electrogravity that is formulated in the weak-gravity limit, and so the vierbein fields  ${e^\mu}_a(x)=\delta^\mu_a + {k^\mu}_a(x) + \cdots$  are expanded, where the first-order fluctuations ${k^\mu}_a(x)$  play the role of spin-1 gauge fields in his gauge gravity theory  \cite{Weinberg_72,Carroll_2004}.  Furthermore, the fundamental metric tensor field $g^{\mu\nu}(x)$ of General Relativity is not a fundamental spin-2 field in the gauge gravity theory.  Instead, the metric tensor field is the bilinear product of two spin-1 vierbein fields, $g^{\mu \nu}(x) = {e^\mu}{}_a(x) e^{\nu a}(x)$, where the  vierbein fields 
are the fundamental fields.  The metric tensor field in the noncoordinate frame is the fixed Minkowski tensor $\eta^{ab}$,  which is transformed into the coordinate frame by contracting it with two vierbein fields,  $g^{\mu \nu}(x)={e^\mu}{}_a(x) {e^\nu}_b(x) \eta^{ab} \mapsto \{{e^\mu}{}_a\gamma^a ,{e^\nu}_b \gamma^b \}/2$ \cite{yepez_arXiv1106.2037_gr_qc}.

A quantum gravity theory can be constructed in the spirit of Einstein's unified field theory of electrodynamics and gravity by generalizing the Minkowski metric. 
The primary ansatz in this quantum gravity theory is the existence of a metric tensor operator. To represent a particle, the classical  Minkowski metric (metric tensor in the noncoordinate basis) is promoted to a metric tensor operator with an antisymmetric (quantum mechanical) part
\begin{align}
\label{Yepez_metric}
  \hat{g}^{ab}   
  &=
     \eta^{ab}
   \mathbb{1}_4 
       +
       \frac{1}{2}[\gamma^a , \gamma^b ] 
  ,
\end{align}
where the  4-vector operator $\gamma^a = ( \gamma^0, \bm{\gamma})$   in the noncoordinate frame has components that are $4 \times 4$ Dirac gamma matrices \cite{1928RSPSA.117..610D,dirac1930principles}.   
 The Dirac matrix for the time-component is unimodular $\gamma_0^2= \mathbb{1}_4$ and hermitian $\gamma_0^\dagger = \gamma_0$, while the  3-vector $\bm{\gamma}=(\gamma_1, \gamma_2, \gamma_3)$ has components that are skew-involution matrices
\begin{align}
\gamma_i^2 = -\mathbb{1}_4
\end{align}
and antihermitian $\gamma_i^\dagger= - \gamma_i$. The Dirac matrices satisfy the anticommutation relations (Clifford algebra)
\begin{align}
\label{Cliiford_algebra}
\{\gamma^a, \gamma^b\} 
=
2 \eta^{ab} \mathbb{1}_4,
\end{align}
where  the  signature of the Minkowski metric tensor is here taken to be $\eta^{ab}=\text{diag}(1,-1,-1,-1)$.  So, inserting (\ref{Cliiford_algebra}) into (\ref{Yepez_metric}), the metric tensor operator in the noncoordinate basis is  expressed in terms the 4-vectors $\gamma^a$  and naturally divided into its manifestly symmetric  (the classical Minkowski metric) and antisymmetric  (the quantum metric) parts 
\begin{align}
\label{noncoordinate_metric_tensor_operator}
  \hat{g}^{ab}   
  &=
       \frac{1}{2}\{ \gamma^a , \gamma^b \}
       +
       \frac{1}{2}[\gamma^a , \gamma^b ] 
       =
       \gamma^a \gamma^b
       .
\end{align}
The 4-vector operator $\gamma^a$ on the righthand side of (\ref{noncoordinate_metric_tensor_operator}) is the ``square root" of the metric tensor operator $\hat{g}^{ab}$ on the lefthand side.  Therefore, the set of 4-vectors $\{\gamma^a\;  |\;  a=0,1,2,3\}$  are immediately identified as the basis 4-vectors of the noncoordinate frame.  In turn, the basis 4-vectors of the coordinate frame, denoted here by $e^\mu(x)$, are  obtained by applying the vierbein fields to effect the transformation from the noncoordinate to the coordinate frame
\begin{align}
\label{coordinate_basis_4_vectors}
\hat  e^\mu(x)
    &=
    {e^\mu}_a(x) \gamma^a
 ,   
\end{align}
 where the frame 4-vector  $e^\mu(x)=(e_0(x), \bm{e}(x))$ 
 is the fundamental field of quantum gravity. 
Finally, the metric tensor operator in the coordinate frame is likewise  obtained by applying the vierbein fields to effect the transformation
\begin{align}
\hspace{-0.1in}
\hat g^{\mu\nu}
&=
 {e^\mu}_a {e^\nu}_b
\hat g^{ab}
\stackrel{(\ref{noncoordinate_metric_tensor_operator})}{=}
    {e^\mu}_a {e^\nu}_b    \gamma^a \gamma^b
\label{Proposition_1_intro_c}
\stackrel{(\ref{coordinate_basis_4_vectors})}{=}
       \hat e^\mu \hat e^\nu
.
 \end{align}

{\it Gauge field theory.}---As a prelude to introducing  quantum gravity as a gauge field theory, consider the  bosonic Lagrangian density (trace-scaled  vierbein gravity) for matter and space 
\begin{align}
\label{vierbein_gravitational_lagrangian_density}
\frac{   {\cal L}}{   \sqrt{-g}}
    =
    \frac{1}{2}
    \nabla^\mu \text{Tr}[{\phi_\circ}    g^{\alpha\beta}]\nabla_\mu \text{Tr}[{\phi_\circ}    g^{\gamma\delta}]
    + 
    R \text{Tr}[{\phi_\circ}    g^{\alpha\beta}]
    .
\end{align}
Variation of the action $\int d^4x\, {\cal L}$ with respect to the vierbein field
 ${e^\mu}_a$
gives the
 Einstein equation  with a predicted cosmological constant term 
\begin{align}
\label{Einstein_equation_expanding_spacetime_with_matter}
G^{\kappa\lambda}
+
\Lambda g^{\kappa\lambda}
=
 \frac{8 \pi G}{c^4}
T^{\kappa\lambda}
,
\end{align}
where the Einstein tensor is $G^{\mu\nu} = R^{\mu \nu}
-
\frac{1}{2} g^{\mu \nu} R$,  the cosmological constant appears in $R^2=\Lambda \text{Tr}[  g^{\alpha\beta}] R$ as the eigenvalue scale, and  the energy momentum tensor is
\begin{align}
 \frac{8 \pi G}{c^4}
  \, T^{\kappa\lambda}   &=
        \left.
        \frac{1}{2\phi }   
         \middle(
           \nabla^\kappa \phi   \nabla^\lambda \phi
         -
        \frac{1}{2}  g^{\kappa\lambda}
        g_{\alpha\beta}\nabla^\alpha\phi \nabla^\beta \phi
                 \right)
                 .
\end{align}
  Here  $\phi
     \equiv
   \text{Tr}(\phi_\circ e^{\mu a} {e^\nu}_a)$ takes the place of the dilaton in Mann's 3+1 dimensional $R=T$ dilaton gravity \cite{Chan_Mann_1994,Scott_Mann_1994}.

The quantum gravity theory  dual to the bosonic theory (\ref{vierbein_gravitational_lagrangian_density}) is the fermionic gauge field theory
\begin{align}
\nonumber
{\cal L}^\text{dual}
& =
\sqrt{-g}\Big[
 i \hbar c \overline{\varepsilon} \hat g^{\mu\nu}\gamma_\mu\left(\partial_{\nu}+\frac{i \mathfrak{e} {A_\nu}^a \hat \Upsilon^{(1)}_a}{\hbar c}\right)\varepsilon
 \\
 &
-
 \frac{1}{4}F_{\mu\nu a}F^{\mu\nu a}
  (\text{Tr}[\phi_\circ g^{\alpha\beta}] -2 \phi_\circ )
 \Big]
\label{quantum_gravity_Lagrangian_density}
,
\end{align}
 where the fermion field operator's equal-time anticommutation relations are  $\{ {\varepsilon}_s(\bm{x}), {\varepsilon}_t(\bm{y})\}
 =0$, $\{ {\varepsilon}_s^\dagger(\bm{x}), {\varepsilon}_t^\dagger(\bm{y})\}=0$ and $
 \{ {\varepsilon}_s(\bm{x}), {\varepsilon}_t^\dagger(\bm{y})\}
 =\delta^{(3)}(\bm{x}-\bm{y})\delta_{st}$, 
   where $s$ and $t$ denote the spinor components of $\varepsilon$. The fermions interact unitarily via a
 gravitational 4-potential   $A^{\mu a} = ({A_0}^a, \bm{A}^a)$, where  $F^{\mu\nu a} =\partial^\mu A^{\nu a} - \partial^\nu A^{\mu a}$ is the gravitational field strength tensor for $a=0,1,2,3$. The spacetime indices are $\mu,\nu = 0,1,2,3$ and 
 $g \equiv -\text{Det}(g_{\rho\sigma})$.  
    The generator for the gravitational gauge group is 
      $\hat \Upsilon^{(1)}_a
      =
      i 4\pi
       \ell
       {e^\beta}_b \partial_a e_{\beta c} S^{bc}$, 
where $\ell$ is the Planck length. The gravitational charge unit is $\mathfrak{e}=\sqrt{\hbar c/(8\pi)}$. 
In the noncoordinate basis, the angular momentum generators in the spin and position representations  of the Lorentz group  are
 $S^{ab} = \frac{1}{4}[\gamma^a, \gamma^b]$ and
 $\hat J^{ab}= i\left(  \gamma^a \ell\partial^b - \gamma^b \ell\partial^a \right)$ (used below). 
${\cal L}^\text{dual}$ is a generalization of flat-space quantum field theory where a metric tensor  operator $\hat g^{\mu\nu}$ 
encodes the fermion's mass and intrinsic spin as torsion in space in the vicinity of the $\varepsilon$ fermion. The dimensionful frame 4-vector is $\hat A^\mu=\mathfrak{e}\hat e^\mu/\ell$.

Varying the action 
$\int dx^4 {\cal L}^\text{dual}$ with respect to $\overline{\varepsilon}$ and $A_{\nu a}$ gives the Euler-Lagrange equations
\begin{subequations}
\label{Yepez_Dirac_Maxwell_London_equations_of_motion_form4}
\begin{align}
    i\hbar c g^{\mu\nu} 
   \hat e_\mu
   \left(
     \partial_\nu 
     + i
     \frac{\mathfrak{e} \hat A_\nu}{\hbar c}
     \right)
     {{\varepsilon}} 
-
       \overline{m} c^2
{{\varepsilon}}
   &  =
     0
\label{Yepez_Dirac_Maxwell_London_equations_of_motion_form4_a}
\\
\label{Yepez_Dirac_Maxwell_London_equations_of_motion_form4_b}
- \frac{1}{(\lambdabar^\text{eff.}_{\text{\tiny L}})^2} 
 A^{\nu a}
&=
\partial_\mu  
F^{\mu\nu a} 
,
\end{align}
\end{subequations}
where $  \overline{m}c^2 \ell \varepsilon=  -( {\hbar c}S^{\mu\nu}\hat J_{\mu\nu}       +
       {3 \mathfrak{e}^2} ) \varepsilon
$ and 
\begin{subequations}
\begin{align}
\label{London_reduced_Compton_wavelength_definition}
 \frac{1}{(\lambdabar^\text{eff.}_{\text{\tiny L}})^2} 
&=
-\frac{1}{\phi-2\phi_\circ}
      \frac{8 \pi \ell^2}{\hbar c}
\left(
     \frac{1}{4}
 {F_{\mu\sigma b}}{F^{\mu\sigma b}}
{\phi}
-
\frac{\hbar c}{8 \pi \ell^2   \lambdabar_\text{\tiny L}^2} 
\right)
\\
\label{initial_London_reduced_Compton_wavelength_definition}
    \frac{1}{\lambdabar_\text{\tiny L}^2}
   & =
           -  
       \sqrt{-g}\,
           \frac{ 4i\hbar c\ell^2}{\mathfrak{e}^2}
      \overline{{{\varepsilon}}}
    \hat e^\mu
\left(           
                 \partial_\mu
+       
\frac{1}{2}
       {e^\beta}_c        
(  \partial_\mu e_{\beta d}) S^{dc}
\right)
    {{\varepsilon}}
    .
\end{align}
\end{subequations}
A gauge-invariant effective  Lagrangian density---that yields  (\ref{Yepez_Dirac_Maxwell_London_equations_of_motion_form4})---is  a superconductive quantum gravity theory
\begin{align}
\nonumber
\frac{{\cal L}^\text{dual}_\text{eff.}
}{\sqrt{-g}}
&=
 i \hbar c \overline{{{\varepsilon}}}  \hat e^\nu
   \left(
     \partial_\nu 
     + i
     \frac{\mathfrak{e} \hat A_\nu}{\hbar c}
     \right)
 {{\varepsilon}}
-
 \overline{m}  c^2
 \overline{{{\varepsilon}}} {{\varepsilon}}
 \\
\label{EG_QFT_Lagrangian_density}
 &
-
 \frac{1}{4}F_{\mu\nu a}F^{\mu\nu a}
 -
 \frac{1}{2(\lambdabar^\text{eff.}_{\text{\tiny L}})^2} 
A_{\nu a}
A^{\nu a}
.
\end{align}
At $t=0$ (massless $\lambdabar^\text{eff.}_{\text{\tiny L}}=\infty$ regime) and at $t \ge t_c$  (high-mass $\lambdabar^\text{eff.}_{\text{\tiny L}}=\lambdabar^c_\text{\tiny L}$ regime), the London relation  \cite{PhysRev.54.947,PhysRev.74.562} is 
\begin{align}
\label{present_day_London_relation_lambdar_star_form}
    \lambdabar_\text{\tiny L}^{\ast}
   &=
      \sqrt{
   \frac{ m_\text{\tiny F}  c^2}{e^2\rho}}
,\quad
\text{or}
\quad
    \lambdabar_\text{\tiny L}^{c \ast}
   =
      \sqrt{
   \frac{ m^c_\text{\tiny F}  c^2}{e^2\rho}}
.
  \end{align}
Hence, the product of the initial scales equals the product of final scales (with superscript $c$) is fixed
\begin{align}
\label{invariant_initial_and_present_mass_scales_products}
    m_\text{\tiny F} m^{\ast}_\text{\tiny L} 
    &=
    m^c_\text{\tiny F} m^{c\ast}_\text{\tiny L} 
    .   
\end{align}
Here $\bullet^\ast$ means the rescaled value of $\bullet$, not the complex conjugate symbol. 
The  fermion mass is $  m^c_\text{\tiny F} = \hbar /( \lambdabar^{c }_\text{\tiny F} c)$  is determined by $\lambdabar^c_\text{\tiny L}$ in  rescaled (\ref{present_day_London_relation_lambdar_star_form}), which can also be  written using  the squared rescaled charge $e^{\ast 2} = 16 \pi G m_\text{\tiny F}^2$
\begin{align}
\label{present_day_London_relation_lambdar_unstarred_form}
    \lambdabar^c_\text{\tiny L} 
   &=
      \sqrt{
   \frac{ (m^c_\text{\tiny F})^2 c^2}{e^{\ast 2}\varrho_\text{\tiny vac}}
   },
\qquad
\text{where}
\qquad
    \lambdabar_\text{\tiny L}^{c \ast} 
    =
    \frac{e^\ast}{e}
      \lambdabar_\text{\tiny L}^c
      .
\end{align}

{\it Nonlinear effective superconductivity.}---The gravitational 4-potential in terms of the 4-frame $\hat e^\mu$ field and the 4-spinor $\varepsilon$ field is 
\begin{align}
\label{gravitational_4_potential_matrix_element}
 {A^\mu}_a(x)
&= 
-\lambdabar_\text{\tiny $L$}^2 \mathfrak{e}
\overline{{{\varepsilon}}}(x)\hat e^\mu(x)   \gamma_a {{\varepsilon}}(x)
.
\end{align}
Inserting this into (\ref{EG_QFT_Lagrangian_density}), the effective  Lagrangian density becomes a nonlinear $(\overline{\varepsilon}\varepsilon)^2$ theory
\begin{align}
\nonumber
\frac{{\cal L}^\text{dual}_\text{eff.}
}{\sqrt{-g}}
&=
 i \hbar c \overline{{{\varepsilon}}}  
 \hat e^\nu
   \left(
     \partial_\nu 
     + i
     \frac{\mathfrak{e} \hat A_\nu}{\hbar c}
     \right)
 {{\varepsilon}}
-
 \overline{m}  c^2
 \overline{{{\varepsilon}}} {{\varepsilon}}
\\
\label{EG_QFT_Lagrangian_density_2}
&
 -
 \frac{\mathfrak{e}^2 \lambdabar_\text{\tiny $L$}^4}{2(\lambdabar^\text{eff.}_{\text{\tiny L}})^2} 
\overline{{{\varepsilon}}}\hat e^\mu   \gamma_a {{\varepsilon}}
\overline{{{\varepsilon}}}\hat e_\mu   \gamma^a {{\varepsilon}}
-
 \frac{1}{4}F_{\mu\nu a}F^{\mu\nu a}
.
\end{align}
Using the 4-frame definition (\ref{coordinate_basis_4_vectors}) and orthonormality ${e_\mu}^a {e^\mu}_b
=
\delta^a_b$, the nonlinear term effective  Lagrangian density reduces to
\begin{align}
\overline{{{\varepsilon}}}\hat e^\mu   \gamma_a {{\varepsilon}}
\,
\overline{{{\varepsilon}}}\hat e_\mu   \gamma^a {{\varepsilon}}
&=
\overline{{{\varepsilon}}}  \gamma_a \gamma_b {{\varepsilon}}
\,
\overline{{{\varepsilon}}}   \gamma^a   \gamma^b {{\varepsilon}}
       \stackrel{(\ref{noncoordinate_metric_tensor_operator})}{=}
\overline{{{\varepsilon}}}  \hat g_{ab} {{\varepsilon}}
\,
\overline{{{\varepsilon}}}   \hat g^{ab} {{\varepsilon}}
.
\end{align}
Moreover, using $\hat{g}^{ab} \stackrel{(\ref{Yepez_metric})}{=} \eta^{ab}\mathbb{1}_4 + 2 S^{ab}$, the nonlinear term has two contributing parts
\begin{subequations}
\begin{align}
\overline{{{\varepsilon}}}  \hat g_{ab} {{\varepsilon}}
\,
\overline{{{\varepsilon}}}   \hat g^{ab} {{\varepsilon}}
&=
\overline{{{\varepsilon}}}  \eta_{ab} {{\varepsilon}}
\,
\overline{{{\varepsilon}}}   \eta^{ab} {{\varepsilon}}
+
4
\overline{{{\varepsilon}}}  S_{ab} {{\varepsilon}}
\,
\overline{{{\varepsilon}}}   S^{ab} {{\varepsilon}}
\\
&=
(\overline{{{\varepsilon}}}  {{\varepsilon}})^2
+
4
\overline{{{\varepsilon}}}  S_{ab} {{\varepsilon}}
\,
\overline{{{\varepsilon}}}   S^{ab} {{\varepsilon}}
.
\end{align}
 \end{subequations}
Therefore, the nonlinear theory may be written as
\begin{align}
\nonumber
\frac{{\cal L}^\text{dual}_\text{eff.}
}{\sqrt{-g}}
&=
 i \hbar c \overline{{{\varepsilon}}}  \hat e^\nu
{\cal D}_\nu
 {{\varepsilon}}
-
 \overline{m}  c^2
 \overline{{{\varepsilon}}} {{\varepsilon}}
 -
 \frac{\mathfrak{e}^2 \lambdabar_\text{\tiny $L$}^4}{2(\lambdabar^\text{eff.}_{\text{\tiny L}})^2} 
(\overline{{{\varepsilon}}}  {{\varepsilon}})^2
\\
\label{EG_QFT_Lagrangian_density_4}
&
-
 \frac{1}{4}F_{\mu\nu a}F^{\mu\nu a}
 -
 \frac{2\mathfrak{e}^2 \lambdabar_\text{\tiny $L$}^4}{(\lambdabar^\text{eff.}_{\text{\tiny L}})^2} 
\overline{{{\varepsilon}}}  S_{ab} {{\varepsilon}}
\,
\overline{{{\varepsilon}}}   S^{ab} {{\varepsilon}}
.
\end{align}

{\it Fundamental mass scales.}---The constant vacuum energy density may be expressed in terms of $m_\text{\tiny F}$ as 
\begin{align}
 \label{vacuum_energy_density_basic_identity_a}
{\cal E}_\text{vac}
=
\varrho_\text{vac} c^2 = 
m_\text{\tiny F}c^2\rho
,
\end{align}
where the number density is
\begin{align}
\label{number_denstiy}
 \qquad
    \rho = \frac{\varrho_\text{vac} }{m_\text{\tiny F}} \sim  
\frac{1}{\frac{4\pi}{3}(\lambdabar_\text{\tiny F})^3}
.
\end{align}
This implies a fermionic mass of
\begin{align}
   m_\text{\tiny F} 
   &=
   \frac{\hbar}{ c}
   \left(\frac{4 \pi  \varrho_\text{vac} c }{3\hbar}\right)^\frac{1}{4}
     \approx
    5.73848 \times 10^{-39}\; \text{kg}
    \approx 0.0032 \;\text{eV}
    ,
\end{align}
using  the observed value of $\varrho_\text{vac}=5.94748 \times 10^{-27} \text{kg}/\text{m}^{3}$  \cite{arXiv:1212.5226v3,arXiv:1303.5062v2}. 

Initially  $S^{\mu\nu}\hat J_{\mu\nu}\varepsilon=0$ when there is no torsion in space,
    so the rest mass energy is
     $\overline{m}c^2 =  -
       {3 \mathfrak{e}^2}/\ell$. 
    Hence, 
    the  torsion-free nonlinear theory (\ref{EG_QFT_Lagrangian_density_4}) takes the simpler form
\begin{align}
\label{EG_QFT_Lagrangian_density_no_twist}
\frac{{\cal L}^\text{dual}_\text{eff.}
}{\sqrt{-g}}
&=
  i \hbar c \overline{{{\varepsilon}}}  \hat e^\nu
{\cal D}_\nu
 {{\varepsilon}}
+
\frac{3 \mathfrak{e}^2}{\ell}
 \overline{{{\varepsilon}}} {{\varepsilon}}
 -
 \frac{\mathfrak{e}^2 \lambdabar_\text{\tiny $L$}^4}{2(\lambdabar^\text{eff.}_{\text{\tiny L}})^2} 
(\overline{{{\varepsilon}}}  {{\varepsilon}})^2
   -
 \frac{1}{4}F_{\mu\nu a}^2
 .
 \end{align}
The essential nonlinear physics is contained in the 
third term. 
The maximum value of $\lambdabar^\text{eff.}_{\text{\tiny L}}$ is infinite, which occurs at $t=0$ when the righthand side of (\ref{London_reduced_Compton_wavelength_definition}) vanishes. 
So the energy contained in the second term on the righthand side of (\ref{EG_QFT_Lagrangian_density_no_twist})   is initially zero, yet it  becomes greater than zero after a critical transition time $t\ge t_c$.  
The highest allowable energy in the second term in (\ref{EG_QFT_Lagrangian_density_no_twist})   bounds 
$\lambdabar^\text{eff.}_{\text{\tiny L}}$ 
to a  finite length. 
The  calculation of the highest energy limit provides a pathway to place a lower bound on the fundamental length scale $\lambdabar^\text{eff.}_{\text{\tiny L}}\ge \lambdabar^\text{min.}_{\text{\tiny L}}$ in the nonlinear superconductive quantum gravity theory. 

Finally, ${\cal L}/\sqrt{-g} 
=           g_{\mu\nu}{\nabla^\mu \phi\nabla^\nu \phi}
  -
           \frac{1}{2}
\left(
-
{2\Lambda}/{\phi_\circ }
\right)
\phi^2
$ equals the Lagrangian density (\ref{vierbein_gravitational_lagrangian_density}),  where $\lambdabar_\text{\tiny L}=\sqrt{-\phi_\circ/(2\Lambda)}$ and where the  vacuum energy density is
$\varrho_\text{vac} c^2
   =
\Lambda { \hbar c}/{(8\pi \ell^2)}$. Hence, the $\phi$'s wavelength   (for  $\phi_\circ = -1$) is
\begin{align}
\lambda_\text{\tiny L}    = 
     \frac{1}{2\ell}\sqrt{-\frac{ \phi_\circ h}{2\varrho_\text{vac} c} }
     \approx
    4.217 \times 10^{26} \; \text{m}
    .
\end{align}

{\it High-pair mass Fermi condensate.}---The  $\varepsilon$  fermion's total mass $\overline{m} = m_\text{\tiny F}$ is a dynamical quantity  and the flux density $\overline{\varepsilon} \varepsilon$ is also a dynamical quantity.
Under  field strengths arising in a high-pair mass Fermi condensate phase, the constant vacuum energy density (\ref{vacuum_energy_density_basic_identity_a}) is 
\begin{align}
\label{S_duality_condition_finite_gauge_field}
{\cal E}_\text{vac}
=
\varrho_\text{vac} c^2 
=
       2m_\text{\tiny F} c^2
    \overline{\varepsilon}
{{\varepsilon}}
=
       2m^c_\text{\tiny F} c^2
   ( \overline{\varepsilon}
{{\varepsilon}})^c
     .
\end{align}
 That is,  in the superconducting Fermi condensate phase, a bosonic pair of $\varepsilon$ fermions has mass $2m_\text{\tiny F}$, while the vacuum energy density  (\ref{vacuum_energy_density_basic_identity_a}) or (\ref{S_duality_condition_finite_gauge_field}) is set by $\overline{m}=m_\text{\tiny F}$.  
Since $\varrho_\text{vac} = 
     m_\text{\tiny F}\rho
$ and the background number density remains set by 
(\ref{number_denstiy}), the flux density in  (\ref{S_duality_condition_finite_gauge_field})  is 
   $\overline{\varepsilon}\varepsilon = 
{\rho}/{2}
=
{\varrho_\text{vac} }/{2m_\text{\tiny F} }$.  
This is a statement of the constancy of the vacuum mass density $ \varrho_\text{vac}$ which is due to the paired $\overline{\varepsilon}\varepsilon$ field. 

So after a high-pair mass Fermi condensate forms (but just before the formation of baryonic matter),   the nonlinear theory (\ref{EG_QFT_Lagrangian_density_no_twist})  stablizes upon achieving rescaled   wavelengths $\lambdabar^c_\text{\tiny F}$ and $\lambdabar^c_\text{\tiny L}$ while satisfying the constraint (\ref{invariant_initial_and_present_mass_scales_products}) as
\begin{subequations}
\label{EG_QFT_Lagrangian_density_no_twist_reconfigured}
\begin{align}
\nonumber
\frac{{\cal L}^\text{dual}_\text{eff.}
}{\sqrt{-g}}
=
 i \hbar c \overline{{{\varepsilon^c}}} \hat e^\nu
   &
   \left(
     \partial_\nu 
     + i
     \frac{e \hat A'_\nu}{\hbar c}
     \right)
 {{\varepsilon}^c}
-
 m^c_\text{\tiny F}  c^2
( \overline{{{\varepsilon}}} {{\varepsilon}})^c
 \\
\label{EG_QFT_Lagrangian_density_no_twist_reconfigured_a}
&-
 \frac{\mathfrak{e}^2 \lambdabar_\text{\tiny $L$}^4}{2(\lambdabar^c_{\text{\tiny L}})^2} 
(\overline{{{\varepsilon}}}  {{\varepsilon})^c}^2
   -
 \frac{1}{4}F'_{\mu\nu a}F'^{\mu\nu a}
 \\
\label{EG_QFT_Lagrangian_density_no_twist_reconfigured_b}
 =
 i \hbar c \overline{{{\varepsilon^c}}}  \hat e^\nu
{\cal D}'_\nu {{\varepsilon^c}}
&-
 m^c_\text{\tiny F}  c^2
( \overline{{{\varepsilon}}} {{\varepsilon}})^c
-
 \frac{1}{2(\lambdabar^c_{\text{\tiny L}})^2} 
{A'}^2_{\nu a}
   -
 \frac{1}{4}{F'}^2_{\mu\nu a}
.
\end{align}
\end{subequations}
Torsion in space associated with the formation of baryonic matter can occur only after the $\varepsilon$ field forms its stable superconducting phase.

The third term in (\ref{EG_QFT_Lagrangian_density_no_twist_reconfigured_b}) is the massive gauge field term.
The maximal energy achievable at a point by an unpaired fermionic field is limited by the Planck energy, $\hbar c/\ell$.  So, the maximal energy achievable by a paired fermionic field is twice the Planck energy, $2\hbar c/\ell$. 
Upon setting the  energy in the third term in (\ref{EG_QFT_Lagrangian_density_no_twist}) to twice the Planck energy gives the following maximal energy density equation using $\overline{{{\varepsilon}}}  {{\varepsilon}}=   {\cal E}_\text{vac}/{
    (2m_\text{\tiny F})}
$: 
\begin{align}
 \label{Planck_energy_upper_bound}
 \frac{2\hbar c}{\ell}
 &=
    \frac{\mathfrak{e}^2 \lambdabar_\text{\tiny $L$}^4}{2(\lambdabar^\text{min.}_{\text{\tiny L }})^2} 
   \frac{{\cal E}_\text{vac}}{
    2m_\text{\tiny F} c^2}
 .
\end{align}
This maximal nonlinearity condition has the solution
\begin{align}
\label{lambdabar_L_min_highest_energy_form}
   \lambdabar^\text{min.}_{\text{\tiny L }}
    &=
    \frac{1}{16}
    \frac{1}{\sqrt{\pi \ell}}
    \frac{\mathfrak{e}^2}{\ell}
    \frac{1}{\sqrt{m_\text{\tiny F}  c^2\varrho_\text{\tiny vac}c^2}}
    \approx
   1.30017 \times 10^{39}
    \;
    \text{m}
    ,
\end{align}
which is rather large for a minimum value and has the form of (\ref{present_day_London_relation_lambdar_unstarred_form}) because of the $\sqrt{\varrho_\text{\tiny vac}}$ in the denominator. 
Therefore, the effective minimum fermion mass $m^\text{min.}_\text{\tiny F}$ of the Fermi condensate can be determined by applying the  London relation 
at the  $\lambdabar^\text{min.}_{\text{\tiny L }}$ scale
\begin{align}
\label{lambdabar_L_min_superconductor_form}
    \lambdabar^\text{min.}_\text{\tiny L} 
   &=
      \sqrt{
   \frac{ (m^\text{min.}_\text{\tiny F})^2 c^2}{e^{\ast 2}\varrho_\text{\tiny vac}}
   }
.
\end{align}
Equating (\ref{lambdabar_L_min_highest_energy_form}) to (\ref{lambdabar_L_min_superconductor_form}) gives a way to  solve for $m^\text{min.}_\text{\tiny F}$  
\begin{subequations}
\begin{align}
  m^\text{min.}_\text{\tiny F} 
   &=
   \frac{1}{16 c^2}
   \sqrt{\frac{e^{\ast 2}}{\pi    m_\text{\tiny F} c^2 \ell}}
   \,
   \frac{\mathfrak{e}^2}{\ell}
   .
\end{align}
\end{subequations}
Finally, the  estimate of a lower bound on $m^\text{eff.}_\text{\tiny F}\ge m^\text{min.}_\text{\tiny F} $ can be expressed in terms of the fundamental constants
\begin{align}
\label{mF_min_bound_1st_estimate}
   m^\text{min.}_\text{\tiny F} 
   &
   =
   \frac{1}{32\pi}
\sqrt{\frac{\hbar}{c \ell} m_\text{\tiny F}}
   =
   \frac{\hbar c}{ 32 \pi \sqrt{\ell} \, c^2}
   \left(\frac{4 \pi  \varrho_\text{vac} c }{
3\hbar}\right)^\frac{1}{8}
      \\
    &\approx
    1.11167 \times 10^{-25}\; \text{kg}
  \approx
  62.36 \text{GeV/c${}^2$}
    .  
\end{align}
This gives a  lower bound estimate of the paired fermion mass  obtained from the nonlinear theory (\ref{EG_QFT_Lagrangian_density_no_twist_reconfigured})
\begin{align}
   m_\text{\tiny pair} 
   &\ge 2 m^\text{min.}_\text{\tiny F} 
   \approx
   124.74 \text{GeV/c${}^2$}
,  
\end{align}
which turns out to be  just under the observed mass of the Higgs boson. A first-order correction is possible. 

{\it Higgs mass formula.}---The background gravitoelectric field---denoted say    $ \bm{E}'_a\gamma^a$ for $a=0,1,2,3$---polarizes the $\overline{\varepsilon}\varepsilon$ paired state that is taken to represent the Higgs boson.  
Thus, the Higgs boson rest mass energy has a gravitational dipole-moment energy contribution which may be  estimated as
  \begin{align}
\label{Eprime_x_dipole_moment} 
- e \bm{E}'_a\gamma^a \cdot \bm{x}
 &\approx
 \frac{\mathfrak{e}^2}{ 4\pi r}
 \quad
 \text{where}
 \quad
  \lambda_\text{\tiny F} = 2\pi r
.
\end{align}
  This induces a small  change to the vacuum energy density in the vicinity of the Higgs boson. To lowest order, this change can be calculated using the dipole approximation   
  \begin{align}
{\cal E}'
&=
     \varrho_\text{vac} c^2
-
 e \bm{E}'_a\gamma^a  \cdot {\bm{x}}\,  
   \frac{\varrho_\text{vac}}{
    m_\text{\tiny F}}
  \stackrel{(\ref{Eprime_x_dipole_moment})}{=}
      \varrho_\text{vac} c^2
      \left(1+\frac{1}{(4\pi)^2}\right)   
.
\end{align}
 Since $\overline{{{\varepsilon}}}  {{\varepsilon}}=   {\cal E}'/{
    2m_\text{\tiny F}}$, the corrected form of (\ref{Planck_energy_upper_bound}) becomes 
\begin{align}
 \frac{2\hbar c}{\ell}
 &=
    \frac{\mathfrak{e}^2 \lambdabar_\text{\tiny $L$}^4}{2(\lambdabar^c_{\text{\tiny L }})^2} 
   \frac{{\cal E}'}{
    2m_\text{\tiny F} c^2}
 .
\end{align}
Finally, using (\ref{present_day_London_relation_lambdar_unstarred_form}) to  eliminate $\lambdabar^c_\text{\tiny $L$}$ gives
 a solution for $m^c_\text{\tiny F}$ 
\begin{align}
   m^c_\text{\tiny F} 
   &
   \approxeq
   \frac{\mathfrak{e}^2}{ 4 \sqrt{\ell} \, c^2}
    \left(1+\frac{1}{(4\pi)^2}\right)^\frac{1}{2}   
   \left(\frac{4 \pi  \varrho_\text{vac} c }{
3\hbar}\right)^\frac{1}{8}
      \\
    &\approx
    1.11518 \times 10^{-25}\; \text{kg}
   \approx
  62.5371 \text{GeV/c${}^2$}
.  
\end{align}
This gives a pair mass  (again twice the fermion mass) of
\begin{align}
   m^c    
   &=
2   m^c_\text{\tiny F} 
   \approx
   2.23036 \times 10^{-25} \text{kg}
\approx
   125.114 \text{GeV/c${}^2$}
   .
\end{align}
Comparing this prediction to the measured value of Higgs mass at the LHC
\begin{align}
   m_\text{\tiny H}^\text{meas.} = 125.10\pm 0.14~\text{GeV}/c^2
   , 
\end{align}
one finds the difference between theory and experiment is only about $0.0143$~GeV$/c{}^2$, which corresponds to about an $0.01139\%$ difference.   The prediction falls right in the middle of the tight experimental error bars, as determined in 2019 by the Particle Data Group \cite{PhysRevD.98.030001}.

{\it Acknowledgements.}---This research was supported by the grant  ``Quantum Computational Mathematics for Efficient Computational Physics"  from the Air Force Office of Scientific Research. 
I would like to thank Prof. Xerxes Tata for helpful discussions.

\end{document}